\documentclass[letterpaper, aps, prl, twocolumn, superscriptaddress, showpacs, nofootinbib]{revtex4}
\pdfoutput=1

\usepackage{graphicx}
\usepackage{bm}
\usepackage{amssymb}
\usepackage{amsmath}
\usepackage{amsfonts}
\usepackage{dcolumn}
\usepackage{natbib}
\usepackage{color}
\usepackage{subfigure}

\usepackage[utf8]{inputenc}

\newcommand{\raiseentry}[1]{\smash{\raise 0.7 em \hbox{#1}}}

\newcommand{\norm}[1]{\lVert #1 \rVert}

\newenvironment{equationarray*}
{\arraycolsep 0.14 em
\begin{eqnarray*}}
{\end{eqnarray*}}

\begin{document}

\title{Formation and Coalescence of Cosmological Supermassive Black Hole Binaries in Supermassive Star Collapse}

\author{C. Reisswig}
\thanks{Einstein Fellow}
\email{reisswig@tapir.caltech.edu}
\affiliation{TAPIR, MC 350-17, California Institute of Technology, 1200 E California Blvd.,
Pasadena, CA 91125, USA}

\author{C. D.\ Ott}
\thanks{Alfred P. Sloan Research Fellow}
\affiliation{TAPIR, MC 350-17, California Institute of Technology, 1200 E California Blvd.,
Pasadena, CA 91125, USA}
\affiliation{Kavli Institute for the Physics and
 Mathematics of the Universe (Kavli IPMU), The University of Tokyo, Kashiwa, Japan}

\author{E. Abdikamalov}
\affiliation{TAPIR, MC 350-17, California Institute of Technology, 1200 E California Blvd.,
Pasadena, CA 91125, USA}

\author{R. Haas}
\affiliation{TAPIR, MC 350-17, California Institute of Technology, 1200 E California Blvd.,
Pasadena, CA 91125, USA}

\author{P. M\"osta}
\affiliation{TAPIR, MC 350-17, California Institute of Technology, 1200 E California Blvd.,
Pasadena, CA 91125, USA}

\author{E. Schnetter}
\affiliation{Perimeter Institute for Theoretical Physics, 31 Caroline
  St.\ N., Waterloo, ON N2L 2Y5, Canada}
\affiliation{Department of Physics, University of Guelph, 50 Stone
  Road East, Guelph, ON N1G 2W1, Canada}
\affiliation{Center for Computation \& Technology, 216 Johnston Hall,
  Louisiana State University, Baton Rouge, LA 70803, USA}

\date{\today}


\begin{abstract}
We study the collapse of rapidly rotating supermassive stars that may have formed in the early Universe.
By self-consistently simulating the dynamics from the onset of collapse using three-dimensional general-relativistic hydrodynamics with fully dynamical spacetime evolution,
we show that seed perturbations in the progenitor can lead to the formation of a system of two high-spin supermassive black holes,
which inspiral and merge under the emission of powerful gravitational radiation that could be observed at redshifts $z\gtrsim10$ with 
the DECIGO or Big Bang Observer gravitational-wave observatories, assuming supermassive stars in the mass range $10^4-10^6\,M_\odot$.
The remnant is rapidly spinning with dimensionless spin $a^*=0.9$. The surrounding accretion disk contains $\sim10\%$ of the initial mass.
\end{abstract}


\maketitle


The observation of supermassive black holes (SMBHs) with masses $M
\gtrsim 10^9 M_\odot $ in the early Universe at redshifts $z\gtrsim7$
(e.g., \cite{mortlock:11}) calls for new pathways for explaining the
existence of such massive black holes when the Universe was less than
$1$ Gyr old (see, e.g.,~\cite{volonteri:12, fan:06} for recent
reviews). Since the time available for the formation of a massive
black hole via accretion from an initial seed black hole is quite
short, it appears likely that the seed objects were quite massive
themselves.  Sufficient initial seed mass could be offered by the
theoretical possibility of supermassive stars (SMSs) with masses $10^4
- 10^6 M_\odot$.  Since SMSs can become general-relativistically unstable
\cite{shapiro:79,baumgarte:99a}, they can collapse and either form
supermassive seed black holes, or, perhaps, die in powerful thermonuclear explosions
with energies $\gtrsim10^{55}\ \rm{ergs}$
\cite{montero:12a,whalen:12,johnson:13}.

A popular scenario for the formation of an SMS is the direct collapse
of a primordial gas cloud in dark matter halos with virial
temperatures $\geq10^4\ \rm{K}$ \cite{oh:02,bromm:03,begelman:06,
  schleicher:10, agarwal:12, choi:13, latif:13a, latif:13b}.  The
required rapid accretion rates \cite{begelman:10} can be maintained by
avoiding early fragmentation, either by isothermal collapse in the
absence of $\rm{H}_2$ cooling
\cite{bromm:03,spaans:06,wise:08,haehnelt:09,latif:11,johnson:11}, or
by turbulent accretion \cite{choi:13, latif:13a, latif:13b}.
Pulsational instabilities or radiative feedback, which could limit the
accretion rate, are sufficiently small to pose no problem for rapid
growth \cite{inayoshi:13, hosokawa:12}.  Since the primordial gas
cloud is likely to carry substantial angular momentum, the centrifugal
barrier must be overcome during the collapse of the halo gas, which
could be achieved by angular momentum transfer via gravitational torques
\cite{choi:13}.

In \cite{shibata:02}, the collapse of a uniformly
rotating SMS was studied in axisymmetry.  Later, \cite{saijo:04,
  saijo:09} investigated the three-dimensional collapse of
differentially rotating SMSs to SMBHs and found that the collapse
proceeds axisymmetrically.
These studies were complemented by \cite{zink:06, zink:07}, who investigated
the collapse of rapidly differentially rotating SMSs with small seed
perturbations that lead to fragmentation during collapse.  These
authors found that a small initial $m=1$ density perturbation grows
exponentially and leads to a single, off-centered fragment that
collapses to a black hole.  In the case of an initial $m=2$ density
perturbation, two orbiting and collapsing fragments form. Before the two fragments
form black holes, however, a run-away collapse at the center of the SMS
occurs, leading to a single black hole at the center of the
star. These studies used a $\Gamma=4/3$ $\Gamma$-law equation of state
(EOS), which is appropriate for SMSs dominated by radiation pressure.

Recently, \cite{montero:12a} investigated the collapse of
SMSs in axisymmetry using a microphysical EOS that models the effects of radiation
and electron-positron pair production. They found that above central temperatures
$\gtrsim 10^9\ \rm{K}$, pair creation effectively reduces the
adiabatic index below $\Gamma=4/3$, thus accelerating the collapse
once these temperatures are reached.

In this Letter, we show that by assuming a $\Gamma$-law EOS with
slightly reduced $\Gamma<4/3$, e.g., because of pair-creation, it is possible to 
form a binary black hole system, provided certain small seed
perturbations are present at the onset of collapse and the star is rapidly differentially rotating.  
The binary black hole system subsequently inspirals and merges under the emission of
powerful gravitational radiation.


\emph{Methods.}  We employ general-relativistic hydrodynamics with
adaptive mesh-refinement provided by the open source
\texttt{EinsteinToolkit} \cite{moesta:13a,et:12} using the
\texttt{Llama} multipatch infrastructure \cite{reisswig:13a,
  pollney:11}, and a modified version of WENO5 reconstruction
\cite{moesta:13a, tchekhovskoy:07}.  We evolve the spacetime geometry with the Z4c
system \cite{hilditch:12}, extract gravitational waves (GWs) at future
null infinity via Cauchy-characteristic extraction
\cite{reisswig:09, reisswig:10a, reisswig:11ccwave, reisswig:11}, and
find apparent horizons using \texttt{AHFinderDirect}
\cite{thornburg:04}.  We employ a $\Gamma$-law EOS and use an
artificial low-density atmosphere ($10^{-10}$ the central density).  If
not stated otherwise, all units are in $c=G=M=K=1$, where $K$ is the
polytropic scale. Since our models can be rescaled to any desired
mass, we report most numbers in units of mass $M$. For convenience, we
give the conversion factors to cgs units in the Appendix.


\begin{table}[t]
\caption{Black hole Christodoulou masses $M_{\rm BH}$ and
  dimensionless Kerr spin parameter $a_{\rm BH}^*$ for each black hole
  in each model.  In model M2G2, we list the parameters of the
  inspiraling two black holes in the first two rows, and the merger
  remnant by the end of the simulation in the third row.  We also
  report the disk mass $M_{\rm disk}$ by the end of the simulation,
  the measured accretion rate $\dot{M}$, and the radiated GW energy
  $E_{\rm GW}$.
For model M2G2, where multiple resolutions are available, we are able
to provide error bars.  Units are in $c=G=M=K=1$, unless otherwise
specified.}
\label{tab:parameters}
\begin{ruledtabular}
\begin{tabular}{llll}
               & M1G1 & M2G1 & M2G2 \\
\hline
BH mass $M_{\rm BH}$ $[M]$  & $5.5$ & $5.8$  & $3.0\pm0.1$    \\
                            & -     & -  & $3.0\pm0.1$    \\
                            & -     & -  & $5.8\pm0.2$    \\
BH spin $a_{\rm BH}^*$        & $0.9$ & $0.9$ & $0.7\pm0.02$  \\
                            & -    & -     & $0.7\pm0.02$  \\
                            & -    & -     & $0.9\pm0.01$  \\
\hline
bar. disk mass $M_{\rm disk}$ $[M]$ & $1.3$ & $1$  & $0.7\pm0.2$ \\
accretion rate $\dot{M}$            & $1.2\times10^{-3}$ & $2\times10^{-4}$ & $6.7\times10^{-5}$ \\
rad. GW energy $E_{\rm GW}$ $[\%]$ & $0.02$ & $0.16$ & $3.71$ \\
\end{tabular}
\end{ruledtabular}
\end{table}

\emph{Initial conditions.}  We consider 3 initial SMS models. Their
initial data are generated via Hachisu's self-consistent field method
\cite{komatsu:89b,stergioulas:95}, which requires as input
the central density $\rho_c$ of the star, and a polar-to-equatorial
axes ratio $r_p/r_e$ between $0$ and $1$ together with a dimensionless
parameter $A$ to set the degree of differential rotation.  All models
are set up as marginally stable $\Gamma=4/3$ polytropes with scale
$K=1$.  In every case, the central density is set to
$\rho_c=3.38\times10^{-6}\,M^{-2}$, the axes ratio is set to
$r_p/r_e=0.24$, and the differential rotation parameter is set to
$A=1/3$.  With these input parameters, the baryonic mass of the star
becomes $M_{*,\rm bar}=7.0527M$, the gravitational mass of the star is
$M_{* \rm,ADM}=7.0037M$, and the angular momentum becomes
$J=52.206\,M^{-2}$, which corresponds to a dimensionless Kerr spin
parameter of $a^*\equiv J/M^2 = 1.0643$.  The resulting equatorial
radius is $r_e=82M$. Due to the rapid rotation, the star has
quasi-toroidal shape (its density maximum is off-center).  The
various models differ by the type of initial perturbation that is
applied and by how the collapse is induced.  In two models, M1G1
and M2G1, the collapse is accelerated by reducing the polytropic scale
$K$ by $10^{-3}$ initially.  During evolution, the two models use an
adiabatic index $\Gamma=4/3$.  In model M2G2, the collapse is
triggered by reducing $\Gamma=4/3$ to $\Gamma=1.33$.  To induce
fragmentation during collapse, we apply a density
perturbation to the initial configuration.  The perturbation is given by
$\rho_{\rm ini}\rightarrow\rho_{\rm ini} \left(1 + A_m r\,\sin(m\phi)\right)$, 
where $m>0$ is an integer, $A_m$ is the perturbation amplitude, $r$ a
cylindrical coordinate radius, and $\phi$ a cylindrical coordinate
angle.  This perturbation offers reasonable overlap with the
corresponding quasinormal modes of the star.  In model M1G1, we apply
a $m=1$ perturbation, and in models M2G1 and M2G2, we apply a $m=2$
perturbation. We use a perturbation amplitude of
$A_m=10^{-3}/r_e\approx1.22\times10^{-5}$.


\emph{Grid setup.}  We use the multiblock grid setup described in
\cite{reisswig:13a}: A central Cartesian grid is surrounded by six
spherical ``inflated-cube'' grids. Their interface is located at radius $R_{\rm s}=240M$, 
and the outer boundary is at $R_{\rm b}=4000M$. 
The central Cartesian grid is
capable of 2:1 mesh refinement. We use $2$ additional finer grids, one
covering the entire star, the other covering the high density
torus. We perform simulations using three resolutions labeled by $r0$,
$r1$, and $r2$.  In our baseline resolution $r1$, the initially finest
level has a resolution of $\Delta x = 0.4M$.  During evolution, we
track each forming fragment with a moving refinement center.  While a
fragment collapses, we progressively add an additional refinement
level every time the fragment's central density increases by another
order of magnitude.  In total, we switch on up to $3$ additional
refinement levels for each fragment until a black hole forms.  The
finest level has a resolution of $\Delta x = 0.05M$.  The outer
spherical grids have radial spacing $\Delta r=3.2M$ and use $N_{\rm
  ang}=21$ cells per patch and angular direction, which corresponds to
an angular resolution of $\simeq4.3^\circ$. The other resolutions $r0$ and
$r2$ have $25\%$ decreased and $25\%$ increased resolution.  See the
Appendix for demonstration of numerical convergence.


\begin{figure}
  \includegraphics[width=1.\linewidth]{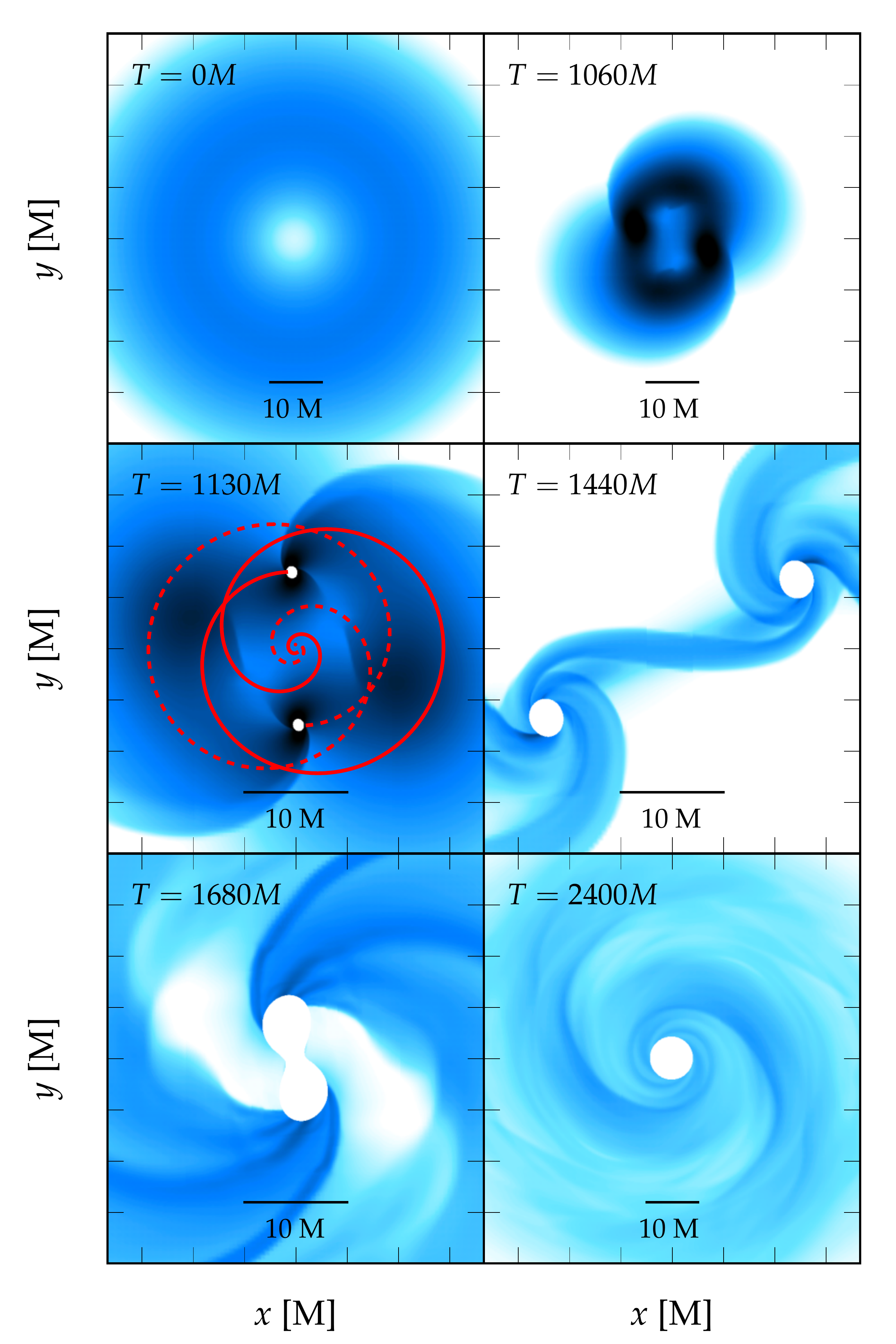}
  \vspace{-0.6cm}
  \caption{Snapshots of a slice of the equatorial density distribution
    of model M2G2. Dark colors indicate high density, light colors
    indicate low density. The logarithmic density colormap ranges from
    $10^{-7}M^{-2}$ (white) to $10^{-3}M^{-2}$ (black). In the bottom
    two panels, the colormap is rescaled to the range $[10^{-8}M^{-2},
      10^{-4}M^{-2}]$ for the sake of presentation. 
      The upper two and the bottom right panels show an
    extent of $\pm40M$, while the remaining panels show an extent of
    $\pm20M$.  }
           \label{fig:m2-dens-frames}
\end{figure}

\begin{figure}
  \includegraphics[width=1.\linewidth]{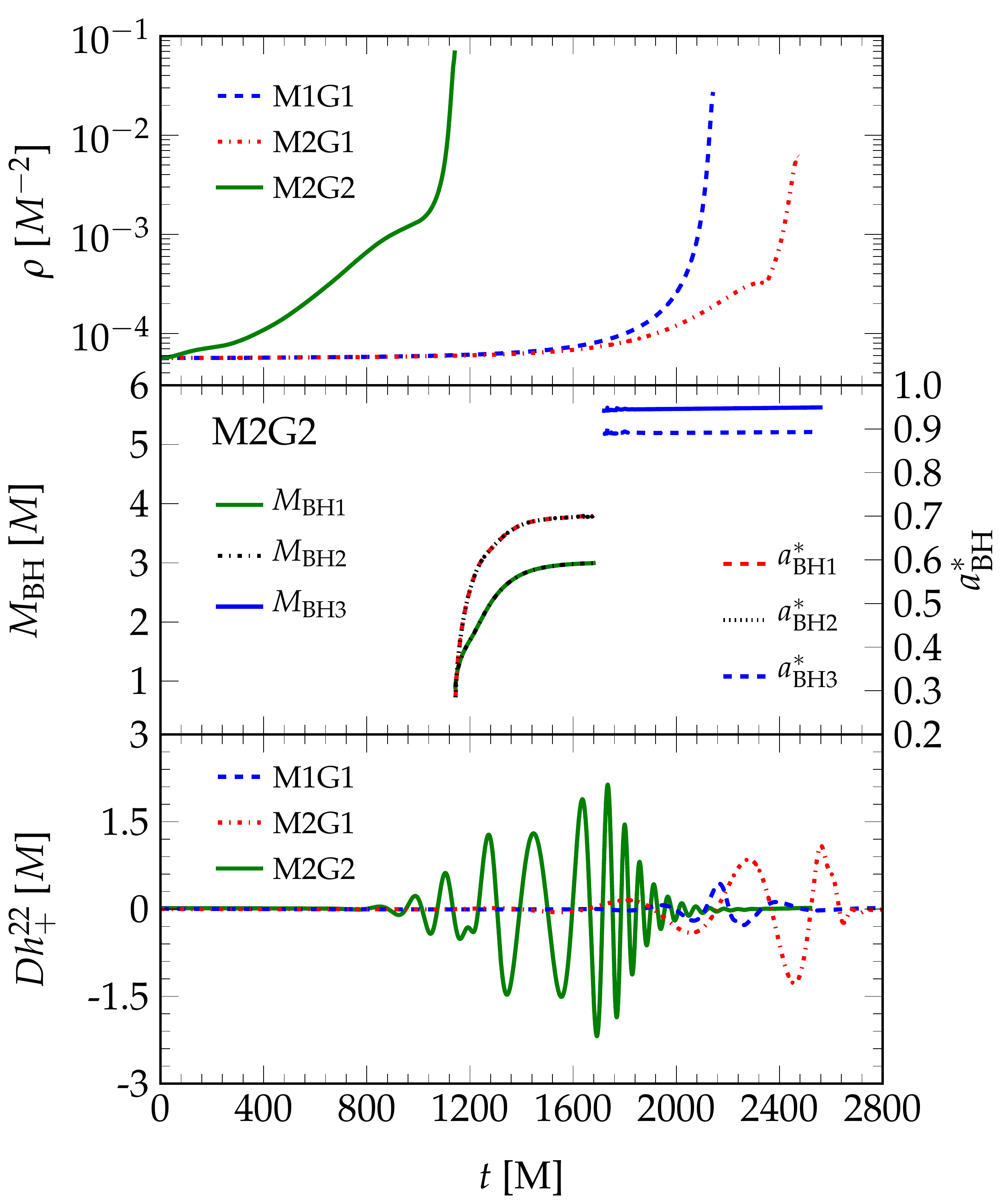}
  \vspace{-0.6cm}
  \caption{\textit{Top panel}: Evolution of the maximum density for all models. 
           \textit{Center panel}: Spin and mass evolution of all three black hole horizons of model M2G2.
           \textit{Lower panel}: $(\ell,m)=(2,2)$ spherical harmonic mode of the ``+'' polarization of 
           the emitted gravitational radiation rescaled by distance $D$.}
           \label{fig:m2-GW-signal}
\end{figure}

\emph{Dynamics.}  Following the initial pressure reduction, either by
reducing $K$ or by reducing $\Gamma$, the rapidly rotating SMSs
undergo accelerated collapse. Depending on the initial density
perturbation, the stars fragment, and collapse accelerates in the
fragments.  Model M1G1, due to its $m=1$ initial density perturbation,
forms a single off-centered fragment which orbits around the star's
center while collapsing. At $T=2145M$, an apparent horizon emerges
around the center of the fragment.
Model M2G1, due to its $m=2$ initial density perturbation, forms two
orbiting fragments that slowly inspiral and collapse. After
$T\simeq2335M$, when the two fragments are still well separated,
run-away collapse occurs at the center of the star, leading to a rapid
increase of the central density.  A single apparent horizon emerges
around the center of the SMS at $T=2470M$.  Models M1G1 and M2G1 were
both considered in \cite{zink:06,zink:07}, and we confirm their
findings, though we are able to continue the evolution well beyond the
appearance of an apparent horizon. We stop the simulation at
$T\simeq3500M$.  Model M2G2 initially follows qualitatively the
evolution of model M2G1. Due to its softer $\Gamma$, however, the two
fragments collapse much faster, forming a pair of two black holes.  In
Fig.~\ref{fig:m2-dens-frames}, we show the equatorial density distribution
of model M2G2 at various stages during the evolution.  
The upper left panel shows the inner
$80M$ of the initial density distribution.  The
toroidal high density ring is clearly visible. The upper right panel
shows a snapshot during collapse. Due to the initial density
perturbation, a strong $m=2$ deformation arises with two inspiraling
high-density fragments.  The center-left panel shows the density
distribution at time $T=1130M$ when an apparent horizon appears within
each of the two still well separated fragments. The apparent horizons
are indicated by white circular regions.  For the purpose of
visualization, they are approximated by the lapse gauge function
$\alpha$ (with the region $\alpha\leq0.19$ closely resembling their
coordinate shapes).  The two nascent black holes have identical mass
and spin. Initially, they have a coordinate separation $D=13.2M$, a
Christodoulou mass $M_{\rm BH}=0.899M$, and dimensionless spin $a_{\rm
  BH}^*=0.286$. Their spin axes are both aligned with the orbital
angular momentum.  Transfer of angular momentum from the
differentially rotating torus onto the black holes via accretion leads
to an outwards spiraling motion until the two black holes reach a
separation $D=24.7M$. From there, they start to inspiral driven by GW
emission and merge after $\sim1.5$ orbits. The tracks of the black
holes are indicated by red (solid and dashed) lines in the center-left panel of
Fig.~\ref{fig:m2-dens-frames}.  In the center-right panel, the black
holes have completed close to one orbit. Around each black hole,
material is dragged and accreted, forming spiral
patterns.
During inspiral, while further accreting material, each of the two
black holes grows to a mass $M_{\rm BH}=3\pm0.1M$ and dimensionless
spin parameter $a_{\rm BH}^*=0.7\pm0.02$ just before merger (center
panel of Fig.~\ref{fig:m2-GW-signal}).  In the bottom left panel of
Fig.~\ref{fig:m2-dens-frames}, the two black holes are about to merge
at time $T=1680M$, and a common apparent horizon appears. 
The bottom-right panel shows the situation at the end of the simulation after the
system has settled to a quasi-stationary state.  The merger remnant
has a Christodoulou mass $M_{\rm BH}=5.8\pm0.2M$, and dimensionless
spin $a_{\rm BH}^*=0.9\pm0.01$.  The surrounding accretion disk
quickly settles to a baryonic mass $M_{\rm disk}=0.7\pm0.2M$.

\emph{GW emission.}
All models emit GWs during collapse and while the fragments form and
orbit around the center of the star. The dominant radiated GW mode for
all models is $(\ell,m)=(2,2)$.  In models with $m=2$ deformation, the
emission is particularly strong. The strongest emission is generated
by model M2G2 due to the two inspiraling and merging SMBHs.  In
Fig.~\ref{fig:m2-GW-signal}, we show the evolution of the maximum
density as well as the $(\ell,m)=(2,2)$ mode of the ``+'' polarization
of the emitted GW signal.  As the maximum density increases, the
oscillatory GW signal rises in amplitude. Following BH formation, a
GW ring-down signal is emitted in models M1G1 and M2G1.  In model M2G2,
following BH formation, we obtain a binary black hole inspiral signal
which increases in frequency and amplitude towards merger. Following
merger, the remnant emits ring-down GWs. The entire GW signal
is in the dHz frequency band for $M_*\sim 10^4\ M_\odot$ SMSs, and in the
mHz band for $M_*\sim 10^6\ M_\odot$ SMSs.

Using modes up to $\ell=8$, we compute the radiated energy emitted in
GWs.  In model M1G1, $0.02\%$ of the initial gravitational mass
$M_{*,\rm ADM}$ is radiated in GWs, $0.16\%$ is radiated in model
M2G1, and $3.71\%$ is radiated in model M2G2 (see
Table~\ref{tab:parameters}).  We compute the detectabilities by the
proposed space-borne eLISA, DECIGO, and Big Bang Observer GW observatories
\cite{eLISA,kawamura:11,cutler:09} over a range of redshifts
$z\in[5,100]$ and source masses $M_*\in[10^4 M_\odot,10^6 M_\odot]$.
Assuming a $\Lambda$CDM cosmology with parameters measured
by the Planck mission \cite{ade:13}, we compute the luminosity
distance $D(z)$ and redshifted mass $(1+z)M_*$ for a given redshift
$z$.  The optimal and angle-averaged signal-to-noise ratios (SNRs)
(e.g.~\cite{flanagan:98a,reisswig:09gw}) can then be obtained from the
Fourier-transformed
angle-dependent GW signal $\tilde{h}(\theta,\phi)$ using the
corresponding theoretical GW observatory sensitivity curves
\cite{eLISA,nishizawa:12,cutler:09}, the luminosity $D(z)$ and
corresponding redshifted mass $(1+z)M_*$.  Assuming a minimum SNR of
$8$ for detection, we obtain a maximum redshift of $z\simeq25$ for a
$M_*=10^4\,M_\odot$ SMS in DECIGO and the Big Bang Observer, provided
the system's spin axis is pointing towards Earth
($(\theta,\phi)=(0,0)$). Both detectors are limited by the white dwarf
confusion noise at low frequencies (e.g.~\cite{nishizawa:12}) and 
have very similar sensitivity for the considered mass range.
Higher mass SMSs are detectable only out to smaller redshifts. For a
$M_*=10^6\,M_\odot$ SMS, we obtain a maximum redshift $z\simeq16$ for
detection.  By assuming a random orientation of the spin axis, we
obtain a mean GW detectability out to $z\simeq23$ for low-mass SMSs,
and $z\simeq13$ for high-mass SMSs.  In eLISA, the signal is barely
detectable at $z\simeq6$, which is outside the relevant range of
redshifts $z\gtrsim10$ at which SMSs are anticipated to exist.

\emph{Discussion.}  We have self-consistently simulated the collapse
of SMSs using $3+1$ general-relativistic hydrodynamics with
approximation-free dynamical spacetime evolution.  We have shown that
it is possible to form a coalescing binary SMBH system in SMS collapse
provided the following conditions are realized: (i) the SMS must be
rapidly differentially rotating, (ii) the gas pressure must be reduced
by a process effectively lowering $\Gamma<4/3$, (iii) a small initial
$m=2$ density perturbation must be present.
The first condition may easily be met since collapsing primordial gas
clouds are likely to carry substantial angular momentum.  If an SMS
forms, it is thus expected to rotate rapidly.  The second condition
is motivated by recent simulations of
SMSs using a microphysical EOS \cite{montero:12a}.  At sufficiently
high central temperatures $\gtrsim10^9\,\rm{K}$ encountered during
the collapse, the effective $\Gamma$ is lowered due to
electron-positron pair-production.  Even if the collapse is triggered
purely gravitationally, the temperature eventually increases, causing
an effective decrease in $\Gamma$.  We find that lowering $\Gamma$ by
not more than $0.25\%$ accelerates the collapse sufficiently to form a
pair of two SMBHs. We note that a reduction by $0.25\%$ is reasonable
given that pair instability, once it sets in, will reduce $\Gamma$ to.
even lower values. Finally, condition (iii) does not seem unlikely,
since recent simulations of the direct collapse of primordial gas
clouds \cite{choi:13} find that central supermassive objects that
may lead to SMSs form $m=2$ structures.
In any case, small seed perturbations are likely to be present at the
onset of collapse.
For our SMS model, the $m=1$ and $m=2$ modes have the fastest and
very similar growth rates~\cite{zink:06,zink:07}.
Therefore, even a tiny $m=2$ bias in an initially randomly perturbed
model may result in fast growth of the $m=2$ mode and formation of two
fragments.  A detailed analysis of this will be subject of future
work.

If $m=2$ fragmentation and binary SMBH formation occurs in SMS
collapse, the coalescence of the SMBH binary will result in a unique
GW signal that can be detected at redshifts $z\gtrsim10$ with
DECIGO and Big Bang Observer for SMSs in the mass range
$10^4-10^6\,M_\odot$. If detected and identified, such a signal would
confirm the existence of SMSs and could potentially inform us about
their rotation and thermodynamics.  Moreover, the interaction between
matter and the binary SMBHs will likely result in an electromagnetic
signature, the details of which will need to be established by future
work.


\acknowledgments We acknowledge helpful discussions with Volker Bromm,
Sterl Phinney, Michele Vallisneri and members of our Simulating
eXtreme Spacetimes (SXS) collaboration
(\url{http://www.black-holes.org}). This research is partially
supported by NSF grant nos.\ PHY-1151197, AST-1212170, PHY-1212460,
and OCI-0905046, by the Alfred P. Sloan Foundation, and by the Sherman
Fairchild Foundation.  CR acknowledges support by NASA through
Einstein Postdoctoral Fellowship grant number PF2-130099 awarded by
the Chandra X-ray center, which is operated by the Smithsonian
Astrophysical Observatory for NASA under contract NAS8-03060.  RH
acknowledges support by the Natural Sciences and Engineering Council
of Canada.  The simulations were performed on the Caltech compute
cluster \emph{Zwicky} (NSF MRI award No.\ PHY-0960291), on
supercomputers of the NSF XSEDE network under computer time allocation
TG-PHY100033, on machines of the Louisiana Optical Network Initiative
under grant loni\_numrel08, and at the National Energy Research
Scientific Computing Center (NERSC), which is supported by the Office
of Science of the US Department of Energy under contract
DE-AC02-05CH11231.


\appendix

\subsection{Numerical Convergence}

\begin{figure}
  \includegraphics[width=1.\linewidth]{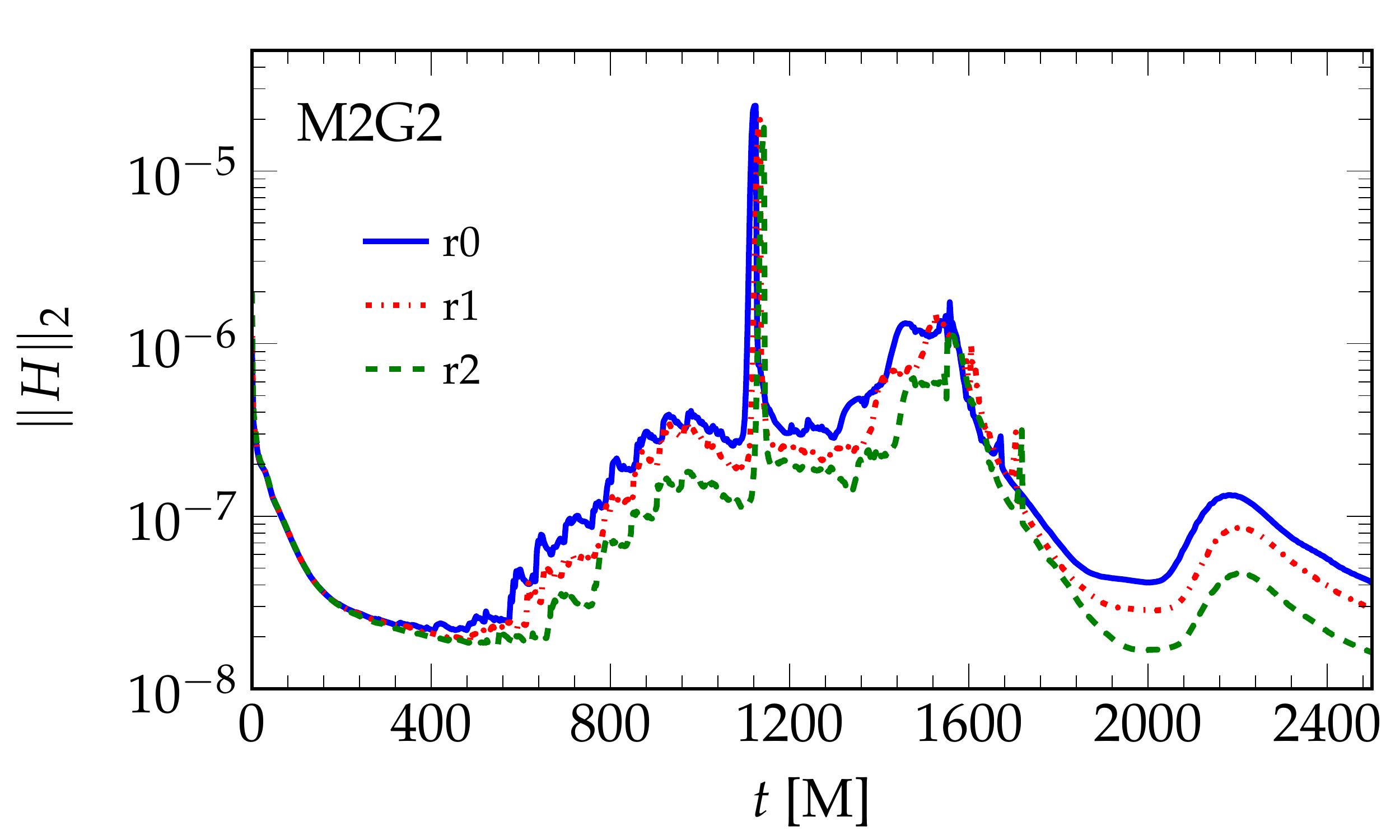}
  \caption{Convergence of the $L_2$-norm of the Hamiltonian constraint
    $\norm{H}_2$ of model M2G2 for the low $r0$, medium $r1$, and high
    resolution $r2$ (not rescaled).  As the resolution is increased,
    the error decreases with a convergence rate between first and
    second-order as expected.  The peaks occur when the black holes
    form, and when the black hole interior is not yet excluded from the 
    calculation of the $L_2$-norm.}
           \label{fig:H_conv}
\end{figure}

We show numerical convergence of our results by considering the
Hamiltonian constraint $H$.  $H = 0$ is one of the four elliptic
constraint equations which arise from the 3+1 ADM decomposition of the
Einstein equations (e.g.~\cite{alcubierre:11}).  Any solution to the
Einstein equation must satisfy these constraints.  At the continuum
level, if they are satisfied initially, they are automatically
satisfied at all times.  During a numerical evolution, however,
numerical error can lead to constraint violations.  As the resolution
is increased, the constraints must converge to zero at the expected
order of accuracy inherent to the numerical scheme.

As detailed in \cite{reisswig:13a}, the overall accuracy of our
simulations is limited by the finite-volume hydrodynamics scheme,
which is second-order accurate in regions where the fluid flow is
smooth, and reduces to first order near shocks and
discontinuities. Thus, we expect between first and second-order
convergence.  We note that the we evolve the spacetime geometry with a
fourth-order finite-difference scheme \cite{reisswig:13a}, but the
overall error is dominated by the fluid evolution.

In Fig.~\ref{fig:H_conv}, we show the $L_2$-norm of the Hamiltonian
constraint $\norm{H}_2$ of model $M2G2$ (which forms a pair of black
holes) for three resolutions labeled by $r0$, $r1$, and $r2$.  The
coarse resolution, $r0$, has $25\%$ reduced resolution compared to
baseline resolution $r1$, and $r2$ has $25\%$ increased resolution.
The $L_2$-norm is taken over the entire domain, excluding the interior
of the black hole horizons.
With increasing resolution, $\norm{H}_2$ decreases with a convergence
rate consistent with first and second order as expected. More
specifically, we find that $\norm{H}_2^{r2}\leqslant 1.25
\norm{H}_2^{r1} \leqslant 1.25\norm{H}_2^{r0}$.  We note that the
numerical evolution is initially non-convergent for the first
$t\leqslant250M$ due to the non-constraint preserving application of
an initial density perturbation.

\subsection{Conversion of Units}

For convenience, we provide the conversion factors from code units
$c=G=M=K=1$ to cgs units in Table~\ref{tab:units}.  As an example for
a $M=10^6M_\odot$ star, a code unit time of $T=1M$ corresponds to a
physical time $t\simeq4.93\,\rm{sec}$, the initial stellar equatorial
radius $r_e=80M$ corresponds to
$\simeq1.2\times10^8\,\rm{km}\simeq0.8\,\rm{AU}$, and the initial
central density $\rho_c=3.38\times10^{-6}M^{-2}$ is
$\simeq2.1\,\rm{g}\, \rm{cm}^{-3}$.

\begin{table}[t]
\caption{Conversion factors for 1 solar mass $M_\odot$ in code units $c=G=M=K=1$
  to cgs units.}
\label{tab:units}
\begin{ruledtabular}
\begin{tabular}{lll}
Quantity & code & cgs \\
\hline
\\ [-2ex]
Mass $M$       & $M$ & $1.9891\times10^{33}(M/M_\odot)$ g \\
Length $L$     & $M$ & $1.4771\times10^5(M/M_\odot)$ cm\\
Time $T$       & $M$  & $4.9271\times10^{-6}(M/M_\odot)$ sec \\
Density $\rho$ & $M^{-2}$ & $6.1716\times10^{17}(M_\odot/M)^2$ g cm${}^{-3}$ \\ 
\end{tabular}
\end{ruledtabular}
\end{table}

\bibliography{bibliography/gw_detector_references,bibliography/cosmology_references,bibliography/bh_formation_references,bibliography/gw_references,bibliography/gr_references,bibliography/sn_theory_references,bibliography/grb_references,bibliography/stellarevolution_references,bibliography/methods_references,bibliography/gw_data_analysis_references,bibliography/NSNS_NSBH_references,bibliography/misc_references,bibliography/cs_hpc_references,bibliography/numrel_references,bibliography/publications-schnetter,bibliography/accretion_disk_references,bibliography/ns_references}

\end{document}